\documentclass[aps,pre,reprint,floatfix]{revtex4-1}
\usepackage{amsmath}
\usepackage{amssymb}
\usepackage{hyperref}
\usepackage{graphicx}
\let\CNI\Theta
\def\Ob{\operatorname{Ob}}
\def\avg#1{\langle#1\rangle}
\def\sgn{\operatorname{sgn}}
\def\SE{\sigma_{\bar x}}
\def\iid{i.i.d.\@}

\def\ErdosRenyi{Erd\H{o}s-R\'enyi }
\begin{document}
\title{A measure for characterizing heavy-tailed networks}
\author{Scott A. Hill}
\email{shill@adrian.edu} 
\affiliation{Adrian College, Adrian MI}
\date{\today} 
\begin{abstract}  
Heavy-tailed networks are often characterized in the literature by their degree distribution's similarity to a power law.  However, many heavy-tailed networks in real life do not have power-law degree distributions, and in many applications the scale-free nature of the network is irrelevant so long as the network possesses hubs.  Here we present the Cooke-Nieboer index (CNI), a non-asymptotic measure of the heavy-tailedness of a network's degree distribution which does not presume a power-law form. The CNI is easy to calculate, and clearly distinguishes between networks with power-law, exponential, and symmetric degree distributions. 

\end{abstract}
\maketitle
\def\needcite#1{{\bf [NEED]}}
\section{Motivation}
The current era of network science research dawned with the discovery  that the relationships in many real-life systems could not be modelled as random graphs\cite{ErdosRenyi}. Instead, real-life networks have hubs\cite{BarabasiAlbert}: nodes with degrees much larger than a random network of the same size and average degree would possess. Their degree distributions are heavy-tailed\cite{Foss}, extending far past the Bernoulli distribution of a \ErdosRenyi network.  

Heavy-tailed networks are usually referred to as ``scale-free networks'' in the literature, which implies that their degree distribution in some ways by a power-law
\begin{equation}\label{powerlaw}
P(x)\sim x^{-\alpha-1},\,\alpha>0.
\end{equation} 
There are a few problems with this.  First, as Broido and Clauset\cite{Broido} point out, the term ``scale-free'' is not always defined in the same way.  Some\cite{Broido,BarabasiAlbert,BarabasiBook,NewmanBook} require that the degree distribution, or at least a portion fo the distribution, follows a strict power-law.  Others use a more lenient definition, like requiring the degree distribution be regularly-varying\cite{Voitalov}, that the distribution be ``well-approximated'' by a power law\cite{BarabasiBook}, or even that the distribution ``looks linear'' on a log-log plot\cite{CSN}.  Some even use the term to describe aspects of a network which are unrelated to its degre distribution, such as the self-similarity of its subgraphs\cite{Song,Krioukov}.  Most of the time, however, when network scientists speak of ``scale-free'' networks, they are really thinking of a network with hubs: that is, a network with a heavy-tailed degree distribution.  This could be dismissed as merely a semantic controversy, but there may be times when the distinction between scale-free and heavy-tailed networks is important.  For example, the proof\cite{Epidemic} that certain scale-free networks have no epidemic threshold depends on the infinite variance of a power-law degree distribution with $\alpha\le2$; networks with finite variance may not share this property.  
 
To determine whether a degree distribution is heavy-tailed, the most common measure is to fit a portion of the distribution to a power-law Eq.~\ref{powerlaw}, no matter its actual shape; the exponent $\alpha$  is known as the \emph{tail index} of the distribution\cite{Cooke,Kotz,Hill,Pickands} and the network.  This asymptotic measure usually depends on a very small fraction of nodes of the network, those residing in the distribution's tail, and philosophically it reinforces the semantic equivalence between heavy-tailed and scale-free networks.

As an alternative, we present a new measure, called the \emph{Cooke-Nieboer index}, which quantifies the heavy-tailedness of a network.  This measure does not presume that the distribution is scale-free, nor is it asymptotic: rather, it is applied to the entire degree distribution rather than just to its tail.  After defining the measure, we will investigate its value for several theoretical distributions and synthetic networks in order to understand its properties.  We will end by applying our measure to real-world networks, comparing it with the tail index $\alpha$ and the ``strength'' of a power-law fit as discussed in \cite{Broido}.

\section{Definition}

\subsection{The Obesity Index}
In the probability literature\cite{Foss}, a distribution $f(x)$ is said to be \emph{heavy-tailed} if

\begin{equation}
\int_{-\infty}^\infty e^{\lambda x}f(x)\,dx=\infty\quad\hbox{for all }\lambda>0.
\end{equation}

Most heavy-tailed distributions of interest fall into a subcategory known as the  \emph{subexponential} distributions, defined as follows\cite{Goldie}: if $X_1,\dots,X_n$ are independent and identically distributed (\iid) random variables chosen from a subexponential distribution, then

\begin{equation}\label{subexponential}
\lim_{x\to\infty}\frac{P(X_1+\dots+X_n>x)}{ P(\max(X_1,\dots,X_n)>x)}=1,\,\hbox{for all }n\ge 1
\end{equation}

In other words, the sum of the random variables is likely to be large if and only if their maximum is likely to be large.  This is the \emph{principle of a single big jump}\cite{Foss}.  (For example, if the cost of cleaning up from natural disasters follows a subexponential distribution, then the total cost of cleanup in any given year is going to be roughly equal to the total cost of the largest disaster that year.)  Power-law distributions and regulary varying distributions\cite{Voitalov} are subsets of the set of subexponential distributions.

To characterize the ``subexponentiality'' of a distribution $X$, Cooke and Nieboer\cite{Cooke} suggest a measure known as the \emph{obesity index}, defined as follows: select four \iid random samples from the distribution and label them in ascending order, so that $X_1\le X_2\le X_3\le X_4$.  Then

\begin{equation}\label{obesity}
\Ob(X)\equiv P(X_4+X_1>X_2+X_3)
\end{equation}

If the distribution is symmetric, then the quantities $X_4+X_1$ and $X_2+X_3$ are equally likely to be larger, and so its obesity index is one-half\cite{Cooke}.  For a subexponential distribution, on the other hand, $X_4$ will probably be larger than the other three variables combined, in which case $X_1+X_4$ must certainly be greater than $X_2+X_3$, and the probability is much greater than one-half.   
The obesity index is a probability, and so ranges from zero to one.  Like skewness and kurtosis, it is independent of offset and positive scaling of the distribution: i.e. 
\begin{equation}\label{independence}
\Ob(aX+b)=\Ob(X),\,a\in {\mathbb R}^+,b\in {\mathbb R}.
\end{equation}
Multiplying the distribution by a negative number reverses the inequality in Eq.~\ref{obesity}, however, so that
\begin{equation}\label{signswitch}
\Ob(b-aX)=1-\Ob(X),\,a\in {\mathbb R}^+,b\in {\mathbb R}.
\end{equation}


\subsection{The Cooke-Nieboer Index}\label{SS:CNI}
For a given distribution $X$, we define the \emph{Cooke-Nieboer index} (CNI) $\CNI(X)$ in a similar way.

{\bf Definition: } Let $X_1,\dots,X_4$ be four \iid random samples chosen from a particular distribution $X$, such as the degree distribution of a network.  We define 
\begin{equation}
\label{cni}
\CNI(X)\equiv E\left\{\sgn\left(\frac12(\max{X_i}+\min{X_i})-\frac14\sum_iX_i\right)\right\},
\end{equation}
where $E\{\cdot\}$ signifies the expectation value and $\sgn(x)$ is the signum function
\begin{equation}
\sgn(x)=\begin{cases} 1, &x>0\\ 0, & x=0\\ -1,&x<0\\\end{cases}.
\end{equation}
For later convenience, we define
\begin{equation}\label{phi}
\Phi(X)\equiv\frac12(\max{X_i}+\min{X_i})-\frac14\sum_iX_i
\end{equation}
so that $\CNI(X)=E\{\sgn(\Phi(X))\}$.

The CNI differs from the obesity index in three ways:
(i)~The CNI ranges from $-1$ to $1$, so that for symmetric distributions, $\CNI=0$; (ii)~it accounts for the finite probability that $X_1+X_4=X_2+X_3$ in discrete distributions; and (iii)~it avoids the term ``obesity'', which may cause confusion in applications of network science to health issues.  Otherwise, for a continuous distribution X, the two measures are simply related:
\begin{equation}\label{conversion}\CNI(X)=2\Ob(X)-1.\end{equation}

The exact CNI can be calculated for a finite distribution with $N$ data points $x_i$, by considering every combination of four points (including duplicates):
\begin{equation}
\CNI=\frac1{N^4}\sum_{i=1}^N\sum_{j=1}^N\sum_{k=1}^N\sum_{l=1}^N\sgn(\Phi(x_i,x_j,x_k,x_l)).
\end{equation}
This naive algorithm runs in $O(N^4)$ time.  If the data points are non-negative integers $x_i=0,1,\dots,M$ and $n_a$ is the number of data points equal to $a$, then we can use the form
\begin{equation}
\CNI=\frac1{N^4}\sum_{a=0}^{M}\sum_{b=0}^{M}\sum_{c=0}^{M}\sum_{d=0}^{M}n_an_bn_cn_d\sgn(\Phi(a,b,c,d))
\end{equation}
instead, which runs in $O(M^4)$ time.

\begin{figure}[t]
\begin{verbatim}
import numpy as np
import random
def cni(degrees,maxerr=1e-3):
   vals=[]
   N=0
   while True:
      four=random.choices(degrees,k=4)
      phi=max(four)+min(four)-0.5*sum(four)
      vals+=[np.sign(phi)]
      N+=1
      sterr=np.std(vals)/np.sqrt(N)
      if(N>20 and sterr<maxerr):
         return np.mean(vals)
\end{verbatim}
\caption{\label{F:code}Sample Python code for calculating the CNI, given a list {\tt degrees} of degrees of the network.  The number 20 in the penultimate line is arbitrary and meant to prevent the code from stopping too soon. The code is written for demonstration purposes and is not particularly efficient; a more sophisticated version can be found in the Appendix.}
\end{figure}

\begin{figure}[htp]
\includegraphics[width=3in]{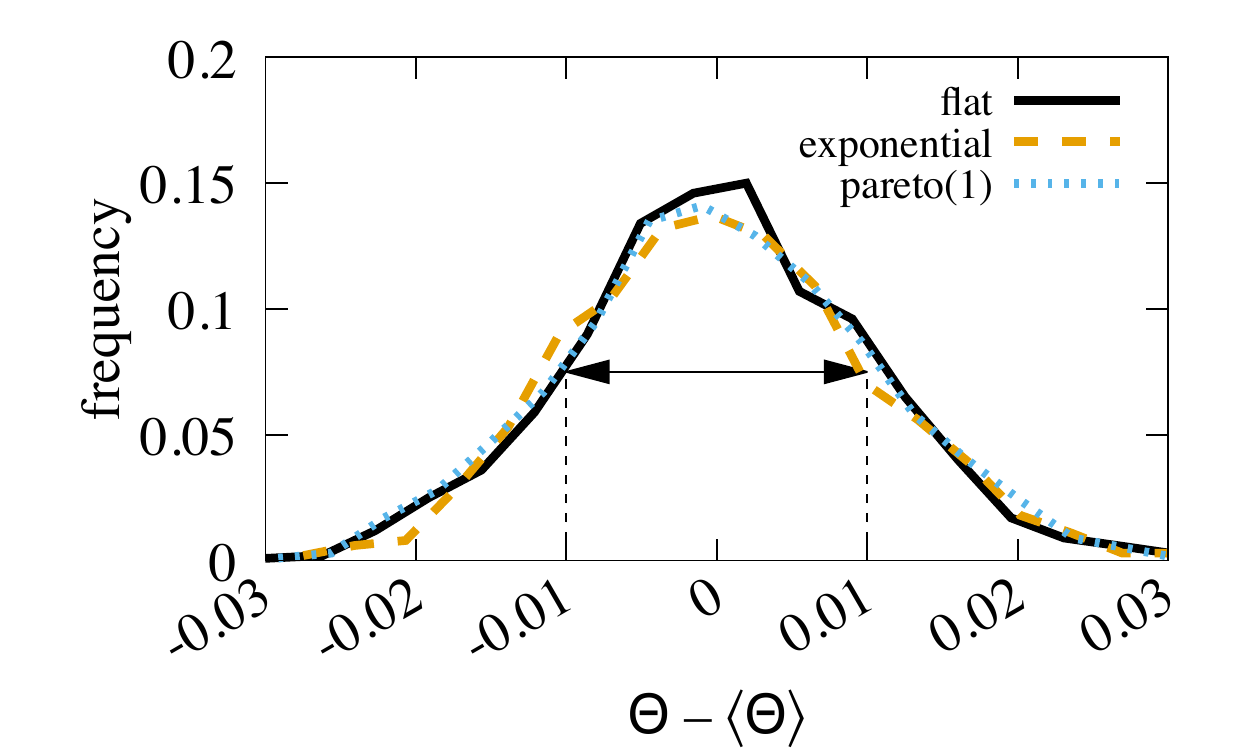}
\caption{\label{F:errspread} A histogram of the deviation from the mean for three distributions: a flat random distribution between 0 and 1, an exponential distribution with $\lambda=1$, and a Pareto distribution with $\alpha=1$.  The CNI was calculated one thousand times using our Monte Carlo algorithm Fig.~\ref{F:code}, each time until reaching a standard error of 0.01.  All three curves are roughly Gaussian with a standard deviation of 0.01, as expected.
}
\end{figure}

\begin{figure}
\includegraphics[width=3in]{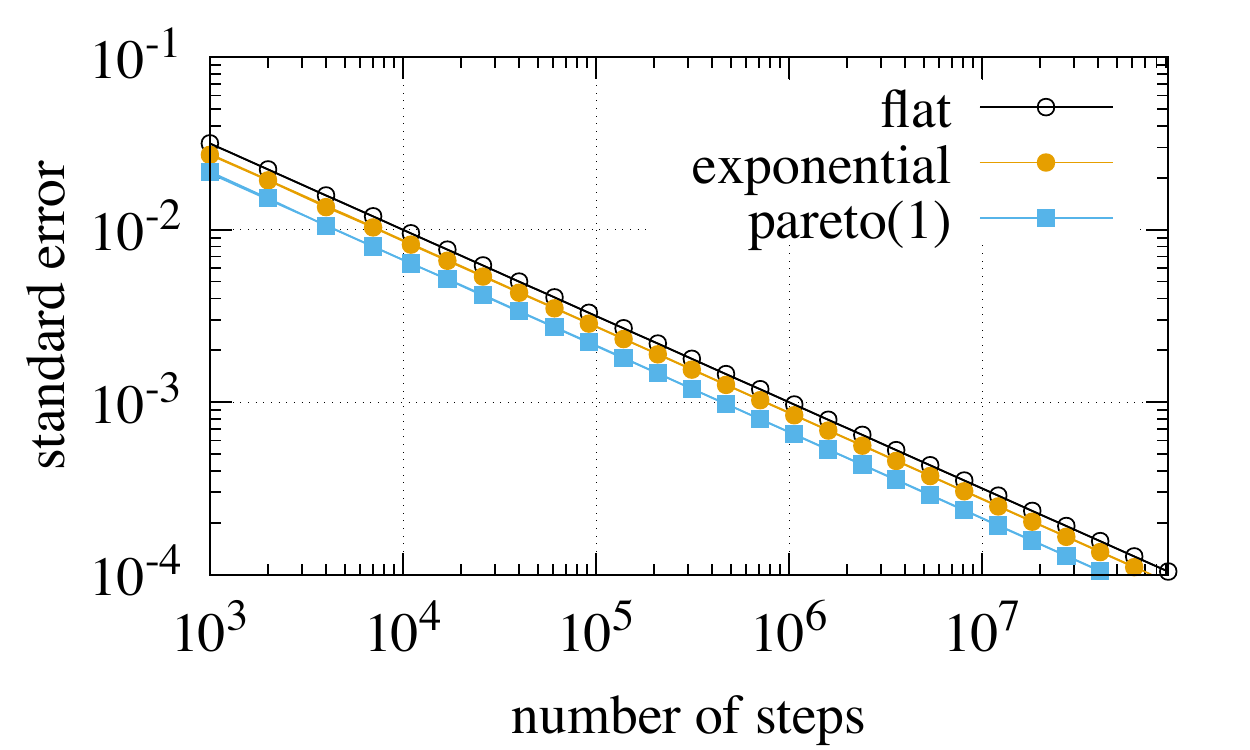}
\caption{\label{F:errsize}
For the same three distributions as in Fig.~\ref{F:errspread}, the number of steps $N$ required to reach a particular standard error $SE$, where a step is a single calculation of $\Phi$ (Eq.~\ref{phi}).  All three curves closely obey the relationship $SE\propto 1/\sqrt{N}$ after one thousand steps.}
\end{figure}

For larger distributions it is sufficient to use a Monte Carlo simulation such as the one in Fig.~\ref{F:code}, calculating $\Phi$ multiple times until some desired standard error $\SE$.  Figure ~\ref{F:errspread} shows that the CNI calculated this way is normally distributed for multiple types of distributions, with a standard deviation equal to $\SE$.  The number of steps required to reach a desired standard error is proportional to $\SE^{-2}$, with a coefficient depending on the type of distribution (Fig.~\ref{F:errsize}).

\section{Distributions}
\begin{figure}[htp]
\includegraphics[width=3in]{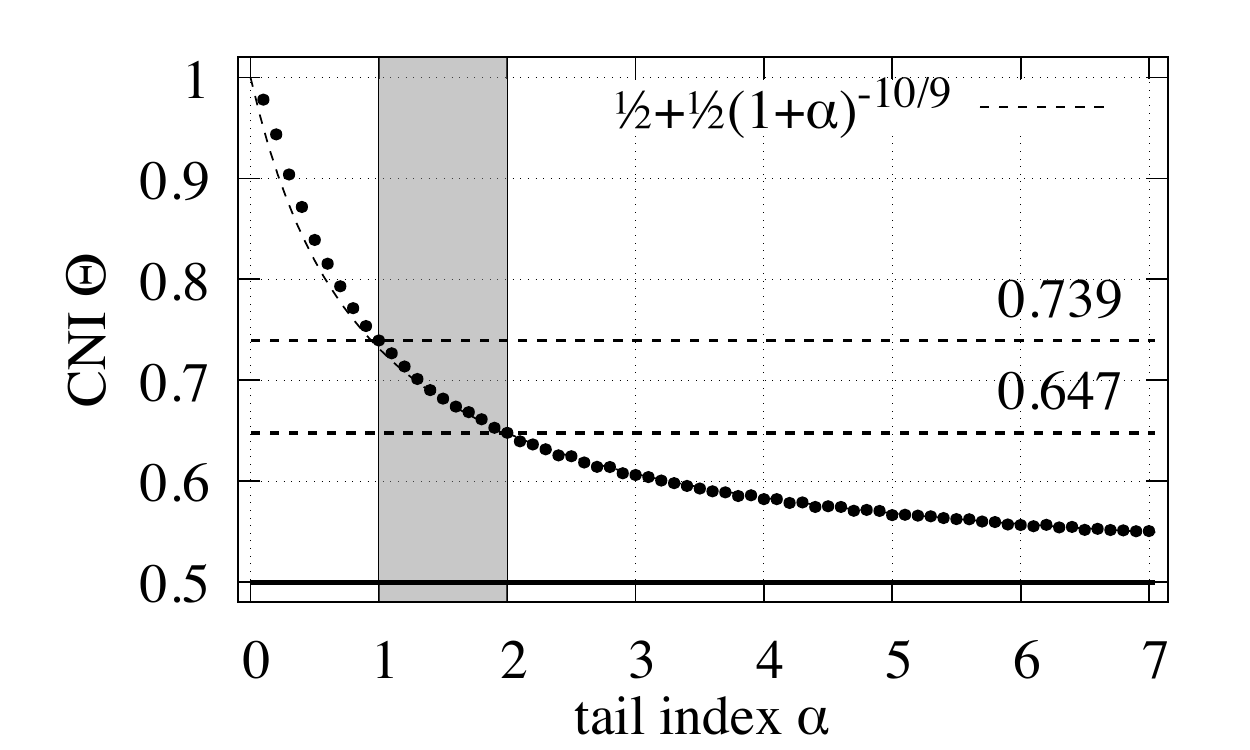}
\caption{\label{F:powerlaw}The CNI of a power-law distribution $1/x^{\alpha+1}$ as a function of its tail index $\alpha$, calculated via numerical simulation.  The grey area highlights the region where most ``scale-free'' networks are found, between $\alpha=1$ and $\alpha=2$\cite{Dorogovtsev,Broido}.  Ref.~\cite{Cooke} calculates the CNI at these values as $2\pi^2-19=0.739$ and $1185-120\pi^2=0.647$, respectively.  There is no closed form for this curve but it is close to the expression $\frac12+\frac12(1+\alpha)^{-10/9}$, which is shown as a dashed line.}
\end{figure}

We saw in Section~\ref{SS:CNI} that $\CNI=0$ for symmetric distributions.  It is shown in \cite{Cooke} that the obesity index that an exponential distribution $P(x)=\lambda e^{-\lambda x}$ has an obesity index is $3/4$ regardless of scale, and thus according to Eq.~\ref{conversion}, $\CNI=1/2$.  Using these two values as boundaries, we divide distributions into three regimes:
\begin{enumerate}
\item {\bf High-CNI} distributions, with $\CNI>0.5$.  These are the \emph{subexponential} distributions, which have heavier tails than the exponential distribution.  They include the power-law distributions, whose CNIs (as shown in Fig.~\ref{F:powerlaw}) range from $\CNI=1$ for $\alpha=0$ to $\CNI=0.5$ as $\alpha\to\infty$.
\item {\bf Low-CNI} distributions, with $0\le \CNI\le 0.5$.  These include the symmetric distributions, the Gumbel distribution\cite{Cooke} $\exp(-e^{-x})$ (with $\CNI\approx 0.25$),  and the binomial and Poisson distributions (as will be seen in Fig.~\ref{F:ER}).
\item {\bf Negative-CNI} distributions, with $\CNI<0$.  These are distributions which have a preponderance of large values and fewer small values: a distribution that grows rather than decays.
\end{enumerate}

\subsection{Bimodal Distribution}\label{SS:bimodal}
To understand how this calculation works, it is useful to consider the simple \emph{bimodal distribution}
\begin{equation}
\label{bimodaldef}
X=\begin{cases}
a&\hbox{with probability $p$}\\
b>a&\hbox{with probability $1-p$}\\
\end{cases}.
\end{equation}
If we choose four samples from this distribution, and $0\le s\le 4$ of them are $a$, it is simple to show that $\Phi$ (Eq.~\ref{phi}) is equal to zero if $s$ is even, $\Phi<0$ if $s=1$, and $\Phi>0$ if $s=3$.  Thus we can calculate the CNI of this distribution precisely:

\begin{equation}\label{bimodalCNI}\begin{split}
\CNI(p)&=\sum_{s=0}^4 {4\choose s}p^s(1-p)^{4-s}\sgn\left(\Phi\right)\\
&=4p^3(1-p)-4p(1-p)^3\\
&=4p(1-p)(2p-1).\\
\end{split} 
\end{equation}
Note that the result does not depend on the values $a$ and $b$.  
\begin{figure}[htp]
\includegraphics[width=3in]{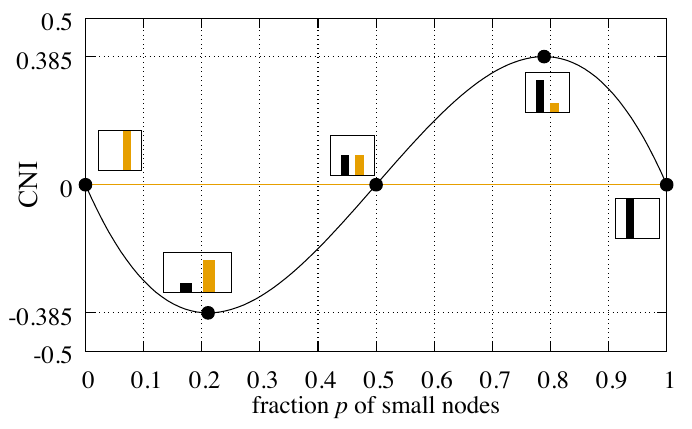}
\caption{\label{F:bimodal}The CNI of the bimodal distribution (Eq.~\ref{bimodaldef}) as a function of $p$.  The small boxes show the relative proportions of the two values ($X=0$ in black, $X=1$ in orange).  The polynomial reaches extreme values of $\pm \frac{2\sqrt3}{9}$ at $p=\frac12\pm\frac{\sqrt3}{6}$.}
\end{figure}

Fig.~\ref{F:bimodal} shows a graph of this polynomial.  Where the distribution is symmetric, at $p=0$, 0.5, and 1, the CNI is zero.  When the smaller values are predominant, as in typical degree distributions, the CNI is positive, with a maximum value of $\CNI=\frac{2\sqrt3}{9}\approx 0.385$ at $p=\frac12+\frac{\sqrt3}{6}\approx 0.79$.  When there are more large values than small values, however, the CNI is negative.

Note that a bimodal distribution can never reach the ``high-CNI'' regime.  By contrast,  \emph{trimodal distributions}
\begin{equation}
\label{trimodaldef}
X=\begin{cases}
a&\hbox{with probability $p$}\\
b>a&\hbox{with probability $q$}\\
c>b&\hbox{with probability $r=1-p-q$}\\
\end{cases}.
\end{equation}
has 
\begin{equation}\label{trimodalCNI}
\CNI=4[p^3(q+r)+q^3(r-p)-(p+q)r^3+3pqr(p-r+jq)]
\end{equation}
where $j=\sgn(c-2b+a)$, \emph{can} reach the high-CNI regime. For example, if $p=2/3$ and $q=r=1/6$, then $\CNI=\frac{14}{27}>\frac{1}{2}$.

\subsection{Changing the Number of Samples}
A natural question to ask regarding our definition is whether there is something about choosing four samples.  In fact, we can generalize Eq.~\ref{phi} to use any number of samples $X_i$:
\begin{equation}
\Phi(X)=\frac12\left(\max X_i+\min X_i\right)-\avg{X_i}.
\end{equation}
The first term $\frac12(\max X_i-\min X_i)$ is the halfway point between the largest and smallest values, and could be thought of as the ``geometric center'' of the samples, while the second term is of course the mean.  When one of the samples is much larger than the others, the mean falls to the negative side of the geometric center, and $\Phi(X)$ is positive.  This makes the CNI a type of skewness measure for the distribution.

\begin{figure}[htp]
\includegraphics[width=3in]{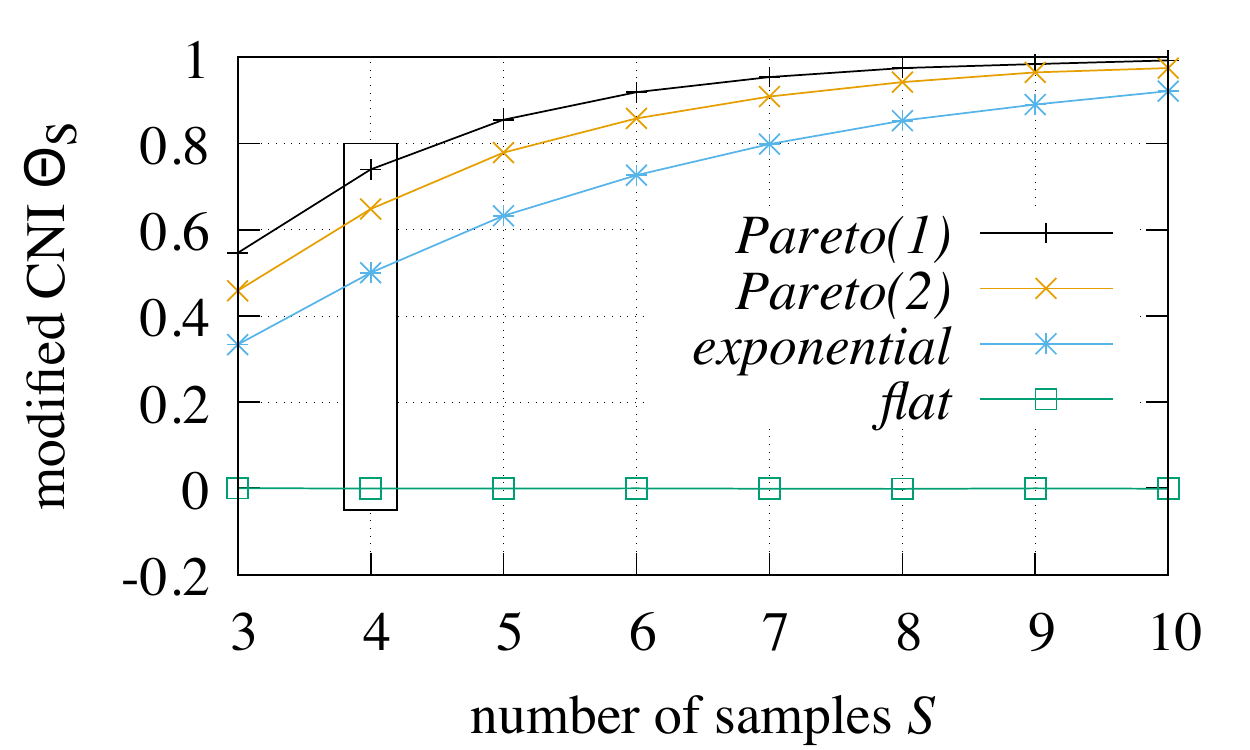}
\caption{\label{F:generalized} The generalized CNI using $S$ samples, for four different random distributions: the Pareto distributions with $\alpha=1$ and $\alpha=2$ ($x^{-2}$ and $x^{-3}$ respectively), the exponential distribution $e^{-\lambda x}$, and a flat random distribution of numbers between 0 and 1. 
}
\end{figure}

Figure~\ref{F:generalized} shows the modified CNI $\CNI_S$ (using $S$ samples) for several basic random distributions.  The value for a flat distribution remains zero throughout, but for others, $\CNI_S$ increases monotonically as the number of samples increases, compressing the ``high-CNI'' regime and expanding the ``low-CNI'' regime.  The value $S=4$ evenly divides the high and low regimes, and so is a reasonable choice for this paper.  Notice that changing the value of $S$ does not change the ordering of these distributions, but this is not true in general.  The generalization of Eq.~\ref{bimodalCNI} for $S$ samples is
\begin{equation}
\CNI_S(p)=\sum_{z=1}^{\lceil S/2-1\rceil}{S\choose z}\left[p^{S-z}(1-p)^z-p^z(1-p)^{S-z}\right]
\end{equation}
and we can show that $\CNI_4(0.77)=0.383$ is less than $\CNI_4(0.79)=0.385$, but $\CNI_7(0.77)=0.732$ is greater than $\CNI_7(0.79)=0.729$.

\section{Networks}\label{S:synthetic}
For an undirected, unweighted network $G$, we define $\CNI(G)$ to be the CNI of its degree distribution; that is, $\CNI(G)=E\left\{\sgn(\Phi)\right\}$ where 
\begin{equation}
\Phi= \frac12(\max k_{n_i}+\min k_{n_i})-\frac14\sum_{i=1}^4{k}_{n_i},\\
\end{equation}
where $n_i\in G$ are nodes and $k_{n_i}$ is the degree of node $n_i$ in $G$.
(For weighted networks, one can let $k_{n_i}$ be the total weight of the edges connected to $n_i$; there is no need for this to be an integer.)

Networks with symmetric degree distributions, such as complete graphs $K_n$ and cycle graphs $C_n$, have $\CNI=0$.  Because $\CNI$ has the same scaling independence as the obesity index (Eq.~\ref{independence}), $\CNI(G\cup G)=\CNI(G)$, although the measure is not otherwise additive. From (Eq.~\ref{signswitch}) it can be shown that 
\begin{equation}\label{converse}
\CNI(\bar G)=-\CNI(G),
\end{equation}
where $\bar G$ is the converse of $G$.

\begin{figure}[htp]
\includegraphics[width=3in]{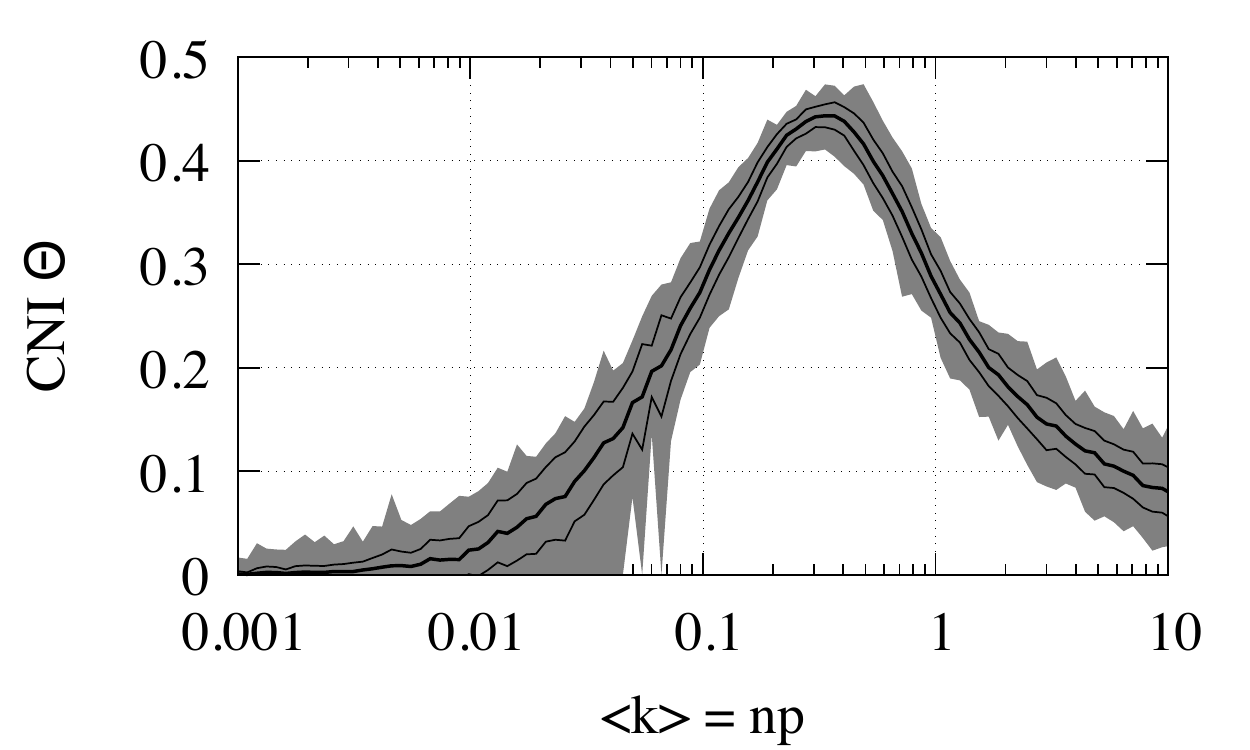}
\caption{\label{F:ER} The CNI $\CNI$ of Erdos-Renyi networks $G(n,p)$ of $n=1000$ nodes with varying average degree $\avg{k}=np$.  One hundred different networks were generated for each value of $\avg{k}$, and their CNI were calculated to a standard error of $0.001$.  The thick central line shows the mean value of $\CNI$; the two lines on either side show one standard deviation away from the mean.  The shaded region shows the range of all values.  Larger values of $N$ result in a similar trajectory but a smaller shaded region.}
\end{figure}

\ErdosRenyi random networks $G(n,p)$ primarily fall in the ``low-CNI regime'' (Fig.~\ref{F:ER}), with the value of $\CNI$ depending strongly on the average degree $\avg{k}=np$ of the network.  The CNI is never negative, but \emph{can} be zero up until a certain threshold ($\avg{k}\approx 0.07$ in the figure), although the average CNI rises steadily with average degree.  The average CNI reaches a maximum value before decreasing until it reaches zero again when $\avg{k}=n-1$.  The significance of the shape of this curve, particularly the threshold where the CNI stops being zero, warrants further study.

\begin{figure}
\includegraphics[width=3in]{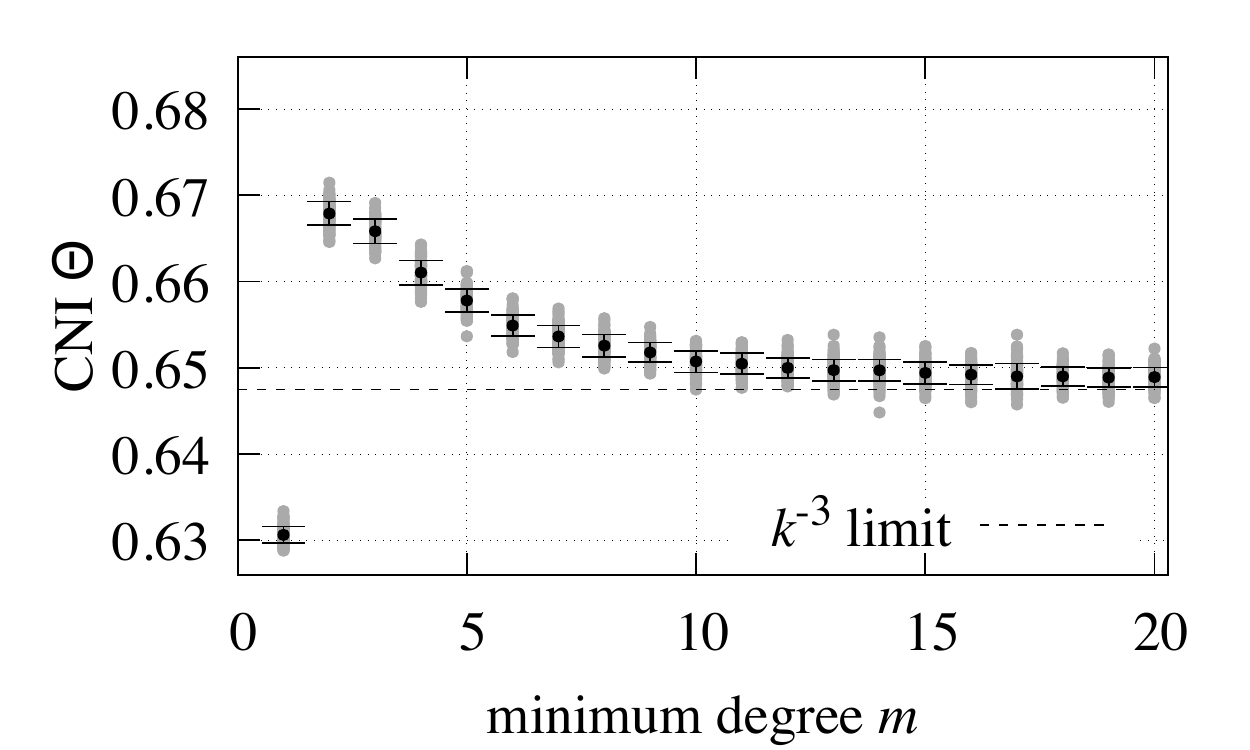}
\caption{\label{F:BA} The CNI for Barab\'asi-Albert networks of $N=$100,000 nodes, as a function of the minimum degree $m$.  The black dot marks the mean value over 100 sampled networks, the error bars show the standard deviation, and the grey dots mark all values.  Note the unusual value at $m=1$.  The dashed line shows the CNI of a power-law distribution $k^{-3}$, which is the value we expect all of these values to converge to\cite{BarabasiAlbert,Bollobas} as $N\to\infty$.}
\end{figure}
Figure~\ref{F:BA} shows that Barab\'asi-Albert networks are high-CNI networks, as is expected, and close to the value measured in Fig.~\ref{F:powerlaw} for a power-law degree distribution with $\alpha=2$.  Notice, however, that the CNI depends on the parameter $m$, which specifies the minimum degree of the network, or alternatively, the number of nodes each new node attaches to when added to the network.  This contradicts \cite{BarabasiAlbert} which says that the infinite-network degree distribution should be $P(k)\propto \frac1{k^3}$ independent of the minimum degree $m$.  This may be a finite size effect, as Barab\'asi-Albert networks are known to converge slowly to their infinite state\cite{Waclaw}.  Recall that the degree distribution is a discrete distribution, unlike the continuous distribution discussed in Fig.~\ref{F:powerlaw}.  The discrepancy may also be due to the non-asymptotic nature of the CNI measure.  According to \cite{Bollobas}, the degree distribution $P(k)$ of such a network should approach
\begin{equation}
\lim_{N\to \infty}P(k)=\frac{2m(m+1)}{k(k+1)(k+2)},\,k\ge m
\end{equation}
For measures that only apply to the tail of the distribution, this can be approximated as $P(k)\propto k^{-3}$; but when the entire distribution is taken to the account, as it is with the CNI, the dependence on $m$ may be more pronounced.  The precise reason for this discrepancy is worthy of further study, as is the jump in value between $m=1$ and $m=2$.

\begin{figure}
\includegraphics[width=3in]{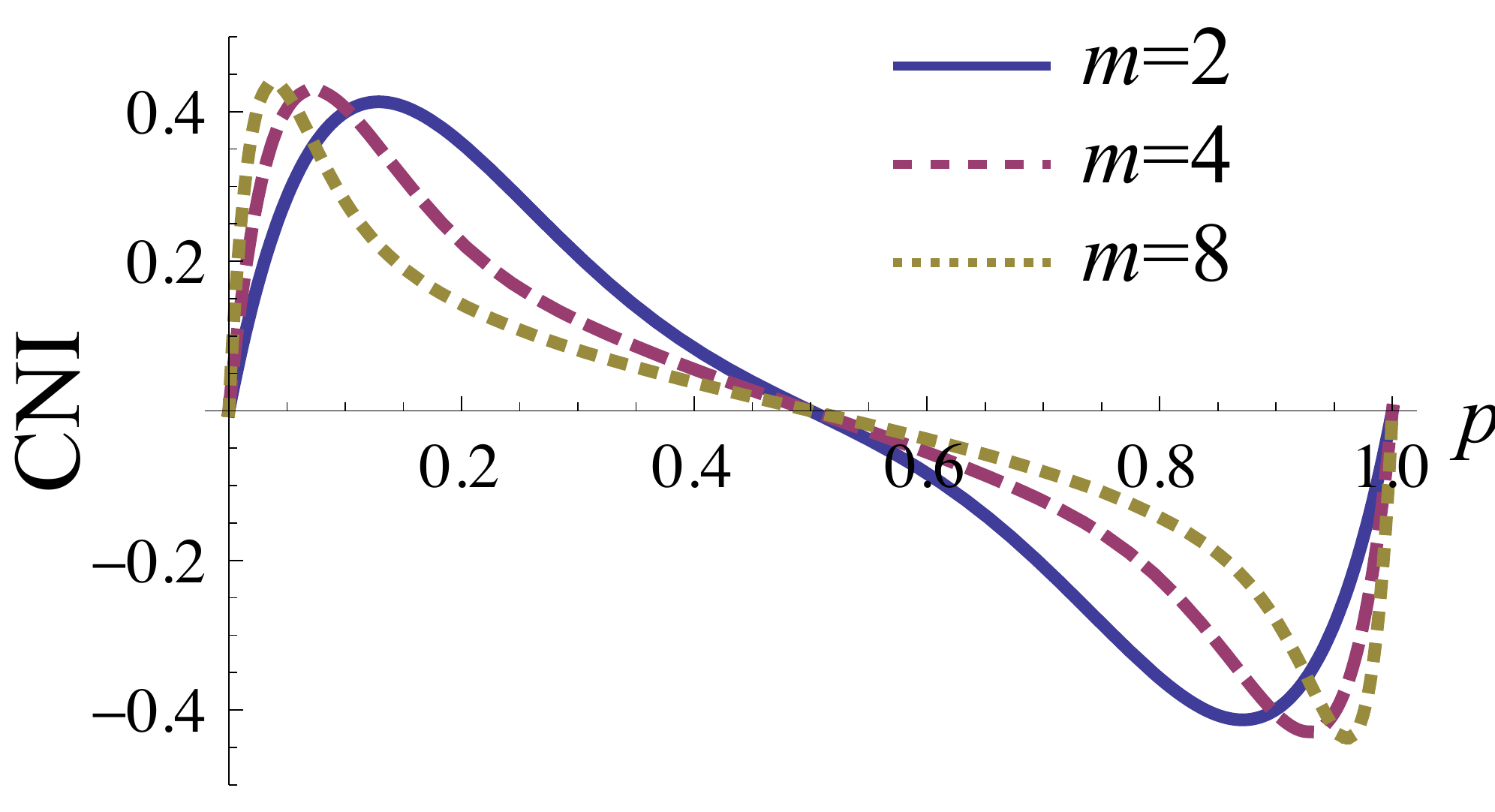}
\caption{\label{F:lattice}The CNI of partial periodic lattices with $m$ nearest neighbors, as a function of edge probability $p$.  If at least half of the edges are kept, then the CNI is negative.}
\end{figure}

Another interesting synthetic network is a \emph{partial periodic lattice} (PPL), in which each node in a lattice with periodic boundary conditions is connected to each of its $m$ nearest neighbors with probability $p$.  For example, a PPL on a square lattice with would have $m=4$. The CNI of a PPL is given by the expression

\begin{multline}
\CNI_{\mathrm{lattice}}(p)=\sum_{i=0}^m\sum_{j=0}^m\sum_{k=0}^m\sum_{l=0}^m \sgn(\Phi(i,j,k,l))\\\times\prod_{s\in\{i,j,k,l\}}{m\choose s}p^s(1-p)^{m-s}.
\end{multline}
and is a $(4m-1)$--degree polynomial.  Figure~\ref{F:lattice} shows this polynomial $\CNI_{\mathrm{lattice}}(p)$ for a few values of $m$.  
Such a network is in the low-CNI regime when $p<0.5$ and there are few well-connected nodes; when $p>0.5$, there are a larger number of high-degree nodes, and $\CNI<0$.  It is a coincidence that the transition between these regimes is equal to the bond percolation threshold of the square lattice\cite{squarepercolation}.

\section{Real-Life Networks}\label{S:real}

We now apply our measure to a set of real-life networks.  We choose to work with the same sample of 927 networks, drawn from the ICON database\cite{ICON}, which are studied in \cite{Broido}.  Following that paper's lead, each non-simple network (i.e. those that are directed, weighted, multipartite, or multiplanar) is used to generate a collection of unweighted, undirected \emph{simple graphs}, according to criteria described in \cite{Broido}.  We define $\bar{\CNI}$ of a \emph{network} to be the median CNI of the network's collection of graphs.

\begin{figure}[htp]
\includegraphics[width=3in]{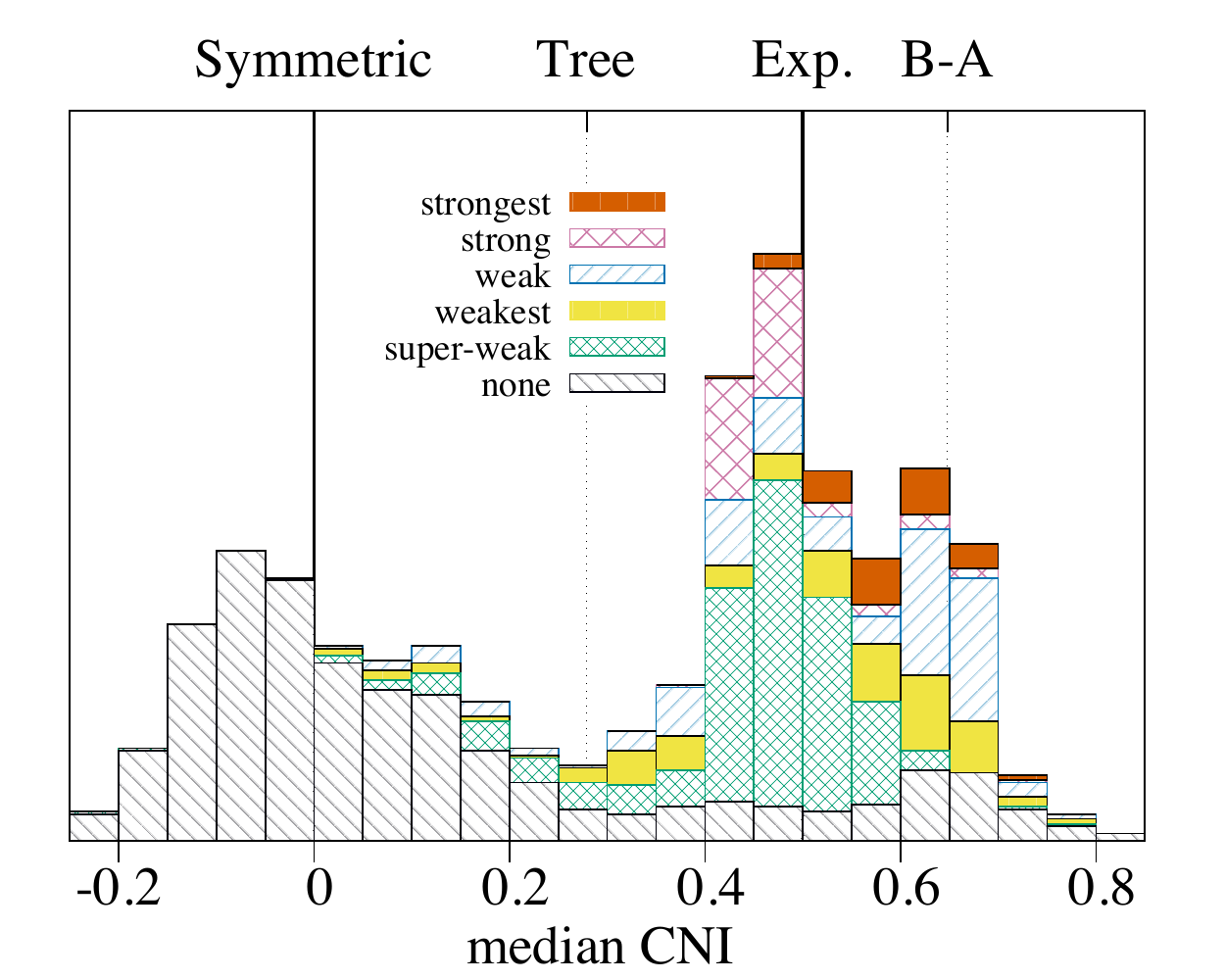}
\caption{\label{F:strength}The distribution of mean CNI for the networks of each strength classification.  Unlike in Ref.~\cite{Broido}, we exclude from the ``super-weak'' category those networks that satisfy the ``weakest'' condition.}
\end{figure}

Figure~\ref{F:strength} shows the distribution of the networks' median CNI, $\bar{\CNI}$.  The average median CNI for all networks is $\avg{\bar \CNI}=0.32\pm 0.27$, but the distribution is bimodal, with one peak around $\CNI=0.5$ and one just below $\CNI=0$.  The negative-CNI peak is made up mostly of planar graphs, specifically United States road networks\cite{USRoads} and fungal growth networks\cite{fungal}; their negative CNI is reminiscent of the partial periodic lattices considered in Section~\ref{S:synthetic}. Excluding these two outlying groups, the average CNI is $\avg{\bar \CNI}=0.49\pm 0.15$, on the boundary between the high- and low-CNI regimes.
Fig.~\ref{F:strength} also breaks the distribution down into the strength classifications used in \cite{Broido}, according to how strong a fit a power-law is to each collection of simple graphs.  Most of the strongest fits to the power-law model have high CNI, though some dip below $0.5$, most significantly the protein-protein interaction network in \emph{Mus musculus}\cite{Musculus} with $\CNI=0.39$.
However, 30\% of networks in the ``weak'' category and below are also high-CNI.  Overall, 31\% of our chosen networks lie in the high-CNI regime; another 24\% are close, in the $0.4\le\CNI<0.5$ range (suggesting a new ``mid-CNI'' regime).  Scale-free networks might be rare, but high-CNI networks are not.

\begin{figure}[htp]
\includegraphics[width=3in]{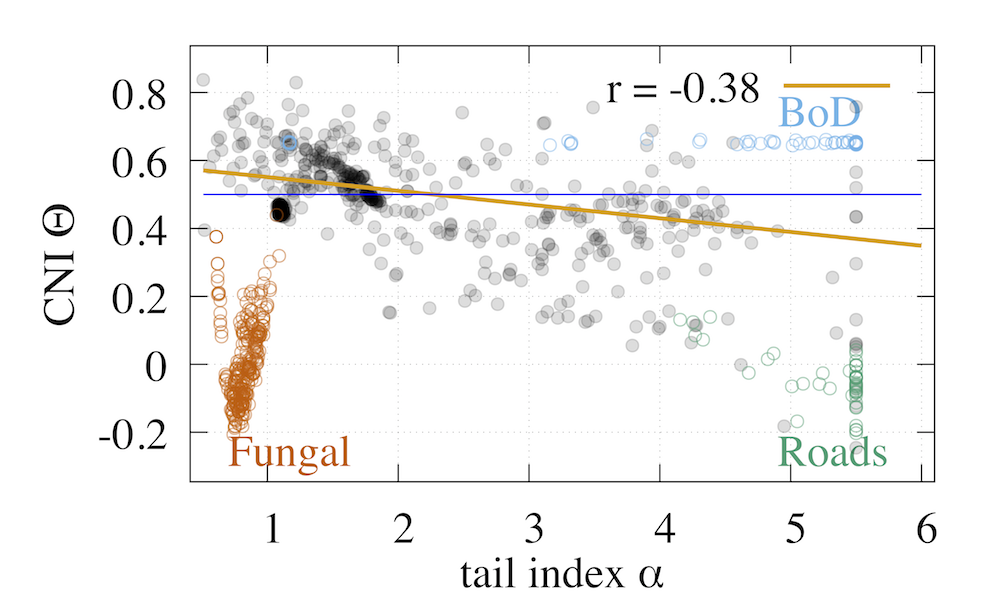}
\caption{\label{F:alpha}The tail index of each simple graph versus its CNI, with linear regression line ($\CNI=-0.04\alpha + 0.59$) showing a moderate negative correlation ($r=-0.38$).  The line crosses the $\CNI=0.5$ subexponential threshold at $\alpha=2.3$.  Three classes of networks are represented with colored open circles: fungal growth networks (red) and US road networks (green) are planar graphs with negative CNI, while the affiliation networks between board directors in Norwegian public limited companies, shown in blue, are further discussed in Fig.~\ref{F:BOD}.}
\end{figure}

Another way to classify the heaviness of a network's tail is with its tail index $\alpha$,  found by fitting the tail of the degree distribution to a power-law $x^{-\alpha-1}$\cite{Broido,CSN,Hill,Pickands}.  Figure~\ref{F:alpha} shows the CNI of each of our simple graphs versus its tail index: the two values have a moderate negative correlation as one might expect, with a Pearson correlation coefficient of $r=-0.38$.  The border between high and low-CNI occurs at $\alpha=2.3$, close to the upper range $\alpha=2$ often cited\cite{Broido,Dorogovtsev} for those networks which are ``scale-free''.

\begin{figure}[htp]
\includegraphics[width=3in]{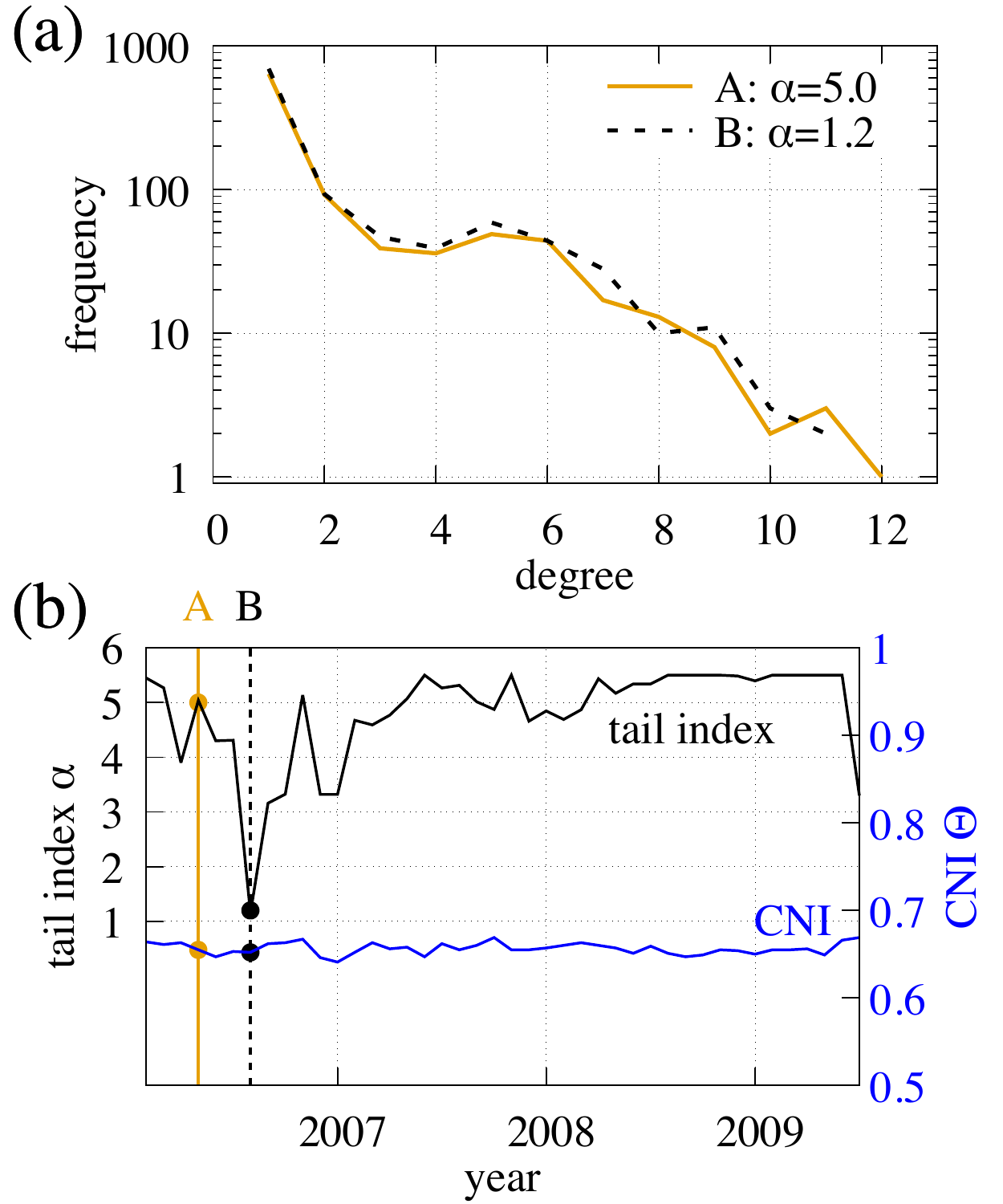}
\caption{\label{F:BOD}
The top graph shows the degree distribution of the network representing the affiliation network between board directors on Norwegian public limited companies\cite{NorwegianBOD} in May 2006 (A) and August 2006 (B).  While having similar degree distributions, their tail indices $\alpha$ are very different ($\alpha=6.0$ and $\alpha=2.2$, respectively).  The bottom graph shows how the tail index and CNI of this network varies over time: while the tail index fluctuates widely, while the CNI remains relatively stable.
}
\end{figure}

However, there are times when the two quantities differ in surprising ways.  Consider the set of affiliation networks between board directors on Norwegian public limited companies\cite{NorwegianBOD}, determined monthly from 2006 through 2009.  These networks have a tail index which varies between 1 and 5.5 (see Fig.~\ref{F:BOD}b), but their CNI is a fairly constant $\CNI=0.656\pm0.007$ throughout.  Do the networks vary significantly or not?  If we look at the degree distributions (Fig.~\ref{F:BOD}a) from two particular months (May 2006 and August 2006) with very different tail indices ($\alpha=5.0$ and $\alpha=1.2$, respectively), we see that the two histograms are quite similar, suggesting that the CNI is a more accurate representation of their heavy-tailed nature.

\section{Conclusion}
We have introduced the Cooke-Nieboer index as a new and potentially useful method for characterizing heavy-tailed networks.  The CNI divides networks into three regimes: high-CNI which includes scale-free networks and other networks with heavy tails, low-CNI which includes random and regular networks, and negative-CNI which includes planar networks which are mostly connected. While presented here in the context of simple graphs, it is easily generalized to apply to weighted and directed networks, 
We have shown (Fig.~\ref{F:alpha}) that our measure is loosely correlated with the tail index of networks, but with certain differences. Philosophically, the CNI avoids the question of whether heavy-tailed networks can be classified as ``scale-free'' or not.  The CNI is also non-asymptotic, but whether this is an improvement on the tail index may depend on the application or one's point of view: the tail index is more sensitive to small changes in the tail, as seen in Fig.~\ref{F:BOD}, but two distributions with the same tail may have considerably different CNIs, depending on the rest of the distribution.  Perhaps the two measures may serve complementary roles, each characterizing certain network behaviors well.   

We hope that this measure will find applications in studies of epidemics, network fragility, and other fields where the distinction between a power-law network and a heavy-tailed network may be important.
There are a number of interesting results in this paper which warrant further study.  The upper limit on the CNI of a bimodal distribution (Fig.~\ref{F:bimodal}) means that a star network, for instance, could never be high-CNI, and there are similar pathological instances of networks which are clearly hub-dominated but which have $\CNI<0.5$.  This could be written off as a mathematical curiosity, but there may be a modification that can address this problem.

The structure of the graph in Fig.~\ref{F:ER}, which shows the distribution of CNI values for \ErdosRenyi networks, has several curious points about it.  Why is there a threshold average degree $\avg{k}$ beyond which one no longer finds networks with $\CNI=0$?  Does the value of $\avg{k}$ that maximizes the average CNI correspond to any other thresholds known to occur in random networks?  

We also saw in Fig.~\ref{F:BA} that the CNI for a Barab\'asi-Albert network depends on the parameter $m$.  The degree distributions of these networks are known to approach a constant power law in the infinite limit independent of the minimum degree, so why is there a steady distinction in the CNI, and why is the CNI so much lower in the $m=1$ case?

In conclusion, we hope that this measure is useful to the network science community at large.

We thank Anne Broido and Aaron Clauset for making their data available in a convenient format at {\tt https://github.com/adbroido/SFAnalysis}; we relied heavily on their data in Section~\ref{S:real}.  We also thank Phil Chodrow and Nicole Eikmeier for useful conversations.   

\section*{Appendix: An Efficient CNI Algorithm}
One can improve the speed of Fig.~\ref{F:code} by implementing a running standard error, such as with Welford's online algorithm\cite{Welford}. However, one can do even better by exploiting the fact that the thing we're taking the average of, $\sgn\Phi$, only takes one of three values.  Suppose we take $N$ sets of four samples from our distribution and calculate $x_i=\sgn\Phi_i$ for each one.  If we define $D\equiv\sum_i x_i$, then the CNI is $\CNI=D/N$.  The variance of this measurement is
$\sigma^2=\frac1{N}\sum_ix_i^2-\avg{x_i}^2$.
Because $x_i^2$ is either zero or one, $\sum_ix_i^2=N-Z$ where $Z$ is the number of times that $\Phi_i=0$.  Thus the variance can be written
\begin{equation}
\sigma^2={N-\frac{Z}{N}}-\left(\frac{D}{N}\right)^2=\frac{N^2-ZN-D^2}{N^2}
\end{equation}
and thus the squared standard error is
\begin{equation}
\SE^2=\frac1{N}\sigma^2=\frac1{N}-\frac{ZN+D^2}{N^3}.
\end{equation}
This confirms the result seen in Fig.~\ref{F:errsize} that $\frac1{\sqrt{N}}$ is an upper-bound and a good approximation for $\SE$, so long as $Z$ and $D$ are both much smaller than $N$.  

The code in Fig.~\ref{F:newcode} uses this insight to calculate the standard error, and calculates the CNI almost 3 times faster than code using the Welford algorithm, and 75 times faster than the code in Fig.~\ref{F:code}.

\begin{figure}[htp]
\caption{\label{F:newcode}A more efficient method of estimating the CNI, written in Python.}
\begin{verbatim}
from random import choices
def cni(degrees, maxerr=0.01):
  S2 = maxerr*maxerr
  Z,D,N = 0,0,0
  while True:
    samp = choices(degrees,k=4)
    val = max(samp)+min(samp)-0.5*sum(samp)
    if val > 0:
      D += 1
    elif val < 0:
      D -= 1
    else:
      Z += 1
    N += 1        
    if not N%20: #only check every 20 steps
      if N*N - Z*N - D*D < S2 * N*N*N:
        return D/N\end{verbatim}
\end{figure}

\bibliography{Hill}

\begin{thebibliography}{26}%
\makeatletter
\providecommand \@ifxundefined [1]{%
 \@ifx{#1\undefined}
}%
\providecommand \@ifnum [1]{%
 \ifnum #1\expandafter \@firstoftwo
 \else \expandafter \@secondoftwo
 \fi
}%
\providecommand \@ifx [1]{%
 \ifx #1\expandafter \@firstoftwo
 \else \expandafter \@secondoftwo
 \fi
}%
\providecommand \natexlab [1]{#1}%
\providecommand \enquote  [1]{``#1''}%
\providecommand \bibnamefont  [1]{#1}%
\providecommand \bibfnamefont [1]{#1}%
\providecommand \citenamefont [1]{#1}%
\providecommand \href@noop [0]{\@secondoftwo}%
\providecommand \href [0]{\begingroup \@sanitize@url \@href}%
\providecommand \@href[1]{\@@startlink{#1}\@@href}%
\providecommand \@@href[1]{\endgroup#1\@@endlink}%
\providecommand \@sanitize@url [0]{\catcode `\\12\catcode `\$12\catcode
  `\&12\catcode `\#12\catcode `\^12\catcode `\_12\catcode `\%12\relax}%
\providecommand \@@startlink[1]{}%
\providecommand \@@endlink[0]{}%
\providecommand \url  [0]{\begingroup\@sanitize@url \@url }%
\providecommand \@url [1]{\endgroup\@href {#1}{\urlprefix }}%
\providecommand \urlprefix  [0]{URL }%
\providecommand \Eprint [0]{\href }%
\providecommand \doibase [0]{http://dx.doi.org/}%
\providecommand \selectlanguage [0]{\@gobble}%
\providecommand \bibinfo  [0]{\@secondoftwo}%
\providecommand \bibfield  [0]{\@secondoftwo}%
\providecommand \translation [1]{[#1]}%
\providecommand \BibitemOpen [0]{}%
\providecommand \bibitemStop [0]{}%
\providecommand \bibitemNoStop [0]{.\EOS\space}%
\providecommand \EOS [0]{\spacefactor3000\relax}%
\providecommand \BibitemShut  [1]{\csname bibitem#1\endcsname}%
\let\auto@bib@innerbib\@empty
\bibitem [{\citenamefont {Erd{\H o}s}\ and\ \citenamefont
  {R{\'e}nyi}(1959)}]{ErdosRenyi}%
  \BibitemOpen
  \bibfield  {author} {\bibinfo {author} {\bibfnamefont {P.}~\bibnamefont
  {Erd{\H o}s}}\ and\ \bibinfo {author} {\bibfnamefont {A.}~\bibnamefont
  {R{\'e}nyi}},\ }\href@noop {} {\bibfield  {journal} {\bibinfo  {journal}
  {Publicationes Mathematicae}\ }\textbf {\bibinfo {volume} {6}},\ \bibinfo
  {pages} {290} (\bibinfo {year} {1959})}\BibitemShut {NoStop}%
\bibitem [{\citenamefont {Barab{\'a}si}\ and\ \citenamefont
  {Albert}(1999)}]{BarabasiAlbert}%
  \BibitemOpen
  \bibfield  {author} {\bibinfo {author} {\bibfnamefont {A.-L.}\ \bibnamefont
  {Barab{\'a}si}}\ and\ \bibinfo {author} {\bibfnamefont {R.}~\bibnamefont
  {Albert}},\ }\href@noop {} {\bibfield  {journal} {\bibinfo  {journal}
  {Science}\ }\textbf {\bibinfo {volume} {286}},\ \bibinfo {pages} {509}
  (\bibinfo {year} {1999})}\BibitemShut {NoStop}%
\bibitem [{\citenamefont {Foss}\ \emph {et~al.}(2013)\citenamefont {Foss},
  \citenamefont {Korshunov},\ and\ \citenamefont {Zachary}}]{Foss}%
  \BibitemOpen
  \bibfield  {author} {\bibinfo {author} {\bibfnamefont {S.}~\bibnamefont
  {Foss}}, \bibinfo {author} {\bibfnamefont {D.}~\bibnamefont {Korshunov}}, \
  and\ \bibinfo {author} {\bibfnamefont {S.}~\bibnamefont {Zachary}},\
  }\href@noop {} {\emph {\bibinfo {title} {An Introduction to Heavy-Tailed and
  Subexponential Distributions}}},\ \bibinfo {edition} {2nd}\ ed.,\ edited by\
  \bibinfo {editor} {\bibfnamefont {T.~V.}\ \bibnamefont {Mikosch}}, \bibinfo
  {editor} {\bibfnamefont {S.~I.}\ \bibnamefont {Resnick}}, \ and\ \bibinfo
  {editor} {\bibfnamefont {S.~M.}\ \bibnamefont {Robinson}},\ Springer Series
  in Operations Research and Financial Engineering\ (\bibinfo  {publisher}
  {Springer},\ \bibinfo {year} {2013})\BibitemShut {NoStop}%
\bibitem [{\citenamefont {{Broido}}\ and\ \citenamefont
  {{Clauset}}(2019)}]{Broido}%
  \BibitemOpen
  \bibfield  {author} {\bibinfo {author} {\bibfnamefont {A.~D.}\ \bibnamefont
  {{Broido}}}\ and\ \bibinfo {author} {\bibfnamefont {A.}~\bibnamefont
  {{Clauset}}},\ }\href@noop {} {\bibfield  {journal} {\bibinfo  {journal}
  {Nature Communications}\ }\textbf {\bibinfo {volume} {10}},\ \bibinfo {pages}
  {1017} (\bibinfo {year} {2019})}\BibitemShut {NoStop}%
\bibitem [{\citenamefont {Barab{\'a}si}(2016)}]{BarabasiBook}%
  \BibitemOpen
  \bibfield  {author} {\bibinfo {author} {\bibfnamefont {A.-L.}\ \bibnamefont
  {Barab{\'a}si}},\ }\href@noop {} {\emph {\bibinfo {title} {Network
  Science}}}\ (\bibinfo  {publisher} {Cambridge University Press},\ \bibinfo
  {year} {2016})\BibitemShut {NoStop}%
\bibitem [{\citenamefont {Newman}(2010)}]{NewmanBook}%
  \BibitemOpen
  \bibfield  {author} {\bibinfo {author} {\bibfnamefont {M.~E.}\ \bibnamefont
  {Newman}},\ }\href@noop {} {\emph {\bibinfo {title} {Networks: An
  Introduction}}},\ \bibinfo {edition} {1st}\ ed.\ (\bibinfo  {publisher}
  {Oxford University Press},\ \bibinfo {year} {2010})\BibitemShut {NoStop}%
\bibitem [{\citenamefont {{Voitalov}}\ \emph {et~al.}(2018)\citenamefont
  {{Voitalov}}, \citenamefont {{van der Hoorn}}, \citenamefont {{van der
  Hofstad}},\ and\ \citenamefont {{Krioukov}}}]{Voitalov}%
  \BibitemOpen
  \bibfield  {author} {\bibinfo {author} {\bibfnamefont {I.}~\bibnamefont
  {{Voitalov}}}, \bibinfo {author} {\bibfnamefont {P.}~\bibnamefont {{van der
  Hoorn}}}, \bibinfo {author} {\bibfnamefont {R.}~\bibnamefont {{van der
  Hofstad}}}, \ and\ \bibinfo {author} {\bibfnamefont {D.}~\bibnamefont
  {{Krioukov}}},\ }\href@noop {} {\bibfield  {journal} {\bibinfo  {journal}
  {arXiv e-prints}\ ,\ \bibinfo {eid} {arXiv:1811.02071}} (\bibinfo {year}
  {2018})},\ \Eprint {http://arxiv.org/abs/1811.02071} {arXiv:1811.02071
  [physics.soc-ph]} \BibitemShut {NoStop}%
\bibitem [{\citenamefont {Clauset}\ \emph {et~al.}(2009)\citenamefont
  {Clauset}, \citenamefont {Shalizi},\ and\ \citenamefont {Newman}}]{CSN}%
  \BibitemOpen
  \bibfield  {author} {\bibinfo {author} {\bibfnamefont {A.}~\bibnamefont
  {Clauset}}, \bibinfo {author} {\bibfnamefont {C.~R.}\ \bibnamefont
  {Shalizi}}, \ and\ \bibinfo {author} {\bibfnamefont {M.~E.}\ \bibnamefont
  {Newman}},\ }\href@noop {} {\bibfield  {journal} {\bibinfo  {journal} {SIAM
  Review}\ }\textbf {\bibinfo {volume} {51}},\ \bibinfo {pages} {661} (\bibinfo
  {year} {2009})}\BibitemShut {NoStop}%
\bibitem [{\citenamefont {Song}\ \emph {et~al.}(2005)\citenamefont {Song},
  \citenamefont {Havlin},\ and\ \citenamefont {Makse}}]{Song}%
  \BibitemOpen
  \bibfield  {author} {\bibinfo {author} {\bibfnamefont {C.}~\bibnamefont
  {Song}}, \bibinfo {author} {\bibfnamefont {S.}~\bibnamefont {Havlin}}, \ and\
  \bibinfo {author} {\bibfnamefont {H.}~\bibnamefont {Makse}},\ }\href@noop {}
  {\bibfield  {journal} {\bibinfo  {journal} {Nature}\ }\textbf {\bibinfo
  {volume} {433}},\ \bibinfo {pages} {392} (\bibinfo {year}
  {2005})}\BibitemShut {NoStop}%
\bibitem [{\citenamefont {{Dmitri Krioukov}}\ and\ \citenamefont
  {Bogu{\~n}{\'a}}(2008)}]{Krioukov}%
  \BibitemOpen
  \bibfield  {author} {\bibinfo {author} {\bibfnamefont {M.}~\bibnamefont
  {{Dmitri Krioukov}}}\ and\ \bibinfo {author} {\bibfnamefont {M.}~\bibnamefont
  {Bogu{\~n}{\'a}}},\ }\href@noop {} {\bibfield  {journal} {\bibinfo  {journal}
  {Physical Review Letters}\ }\textbf {\bibinfo {volume} {100}},\ \bibinfo
  {pages} {078701} (\bibinfo {year} {2008})}\BibitemShut {NoStop}%
\bibitem [{\citenamefont {Pastor-Satorras}\ and\ \citenamefont
  {Vespignani}(2001)}]{Epidemic}%
  \BibitemOpen
  \bibfield  {author} {\bibinfo {author} {\bibfnamefont {R.}~\bibnamefont
  {Pastor-Satorras}}\ and\ \bibinfo {author} {\bibfnamefont {A.}~\bibnamefont
  {Vespignani}},\ }\href@noop {} {\bibfield  {journal} {\bibinfo  {journal}
  {Physical Review Letters}\ }\textbf {\bibinfo {volume} {86}},\ \bibinfo
  {pages} {3200} (\bibinfo {year} {2001})}\BibitemShut {NoStop}%
\bibitem [{\citenamefont {Cooke}\ \emph {et~al.}(2014)\citenamefont {Cooke},
  \citenamefont {Nieboer},\ and\ \citenamefont {Misiewicz}}]{Cooke}%
  \BibitemOpen
  \bibfield  {author} {\bibinfo {author} {\bibfnamefont {R.~M.}\ \bibnamefont
  {Cooke}}, \bibinfo {author} {\bibfnamefont {D.}~\bibnamefont {Nieboer}}, \
  and\ \bibinfo {author} {\bibfnamefont {J.}~\bibnamefont {Misiewicz}},\
  }\href@noop {} {\emph {\bibinfo {title} {Fat-Tailed Distributions: Data,
  Diagnostics, and Dependence}}},\ \bibinfo {series} {Mathematical Models and
  Methods in Reliability Set}\ No.~\bibinfo {number} {1}\ (\bibinfo
  {publisher} {Wiley-ISTE},\ \bibinfo {year} {2014})\BibitemShut {NoStop}%
\bibitem [{\citenamefont {Kotz}\ and\ \citenamefont {Nadarajah}(2000)}]{Kotz}%
  \BibitemOpen
  \bibfield  {author} {\bibinfo {author} {\bibfnamefont {S.}~\bibnamefont
  {Kotz}}\ and\ \bibinfo {author} {\bibfnamefont {S.}~\bibnamefont
  {Nadarajah}},\ }\href@noop {} {\emph {\bibinfo {title} {Extreme Value
  Distributions: Theory and Applications}}}\ (\bibinfo  {publisher} {Imperial
  College Press},\ \bibinfo {address} {London},\ \bibinfo {year}
  {2000})\BibitemShut {NoStop}%
\bibitem [{\citenamefont {Hill}(1975)}]{Hill}%
  \BibitemOpen
  \bibfield  {author} {\bibinfo {author} {\bibfnamefont {B.~M.}\ \bibnamefont
  {Hill}},\ }\href@noop {} {\bibfield  {journal} {\bibinfo  {journal} {The
  Annals of Statistics}\ }\textbf {\bibinfo {volume} {3}},\ \bibinfo {pages}
  {1163} (\bibinfo {year} {1975})}\BibitemShut {NoStop}%
\bibitem [{\citenamefont {{James {Pickands} III}}(1975)}]{Pickands}%
  \BibitemOpen
  \bibfield  {author} {\bibinfo {author} {\bibnamefont {{James {Pickands}
  III}}},\ }\href@noop {} {\bibfield  {journal} {\bibinfo  {journal} {The
  Annals of Statistics}\ }\textbf {\bibinfo {volume} {3}},\ \bibinfo {pages}
  {119} (\bibinfo {year} {1975})}\BibitemShut {NoStop}%
\bibitem [{\citenamefont {Goldie}\ and\ \citenamefont
  {Kl{\"u}ppelberg}(1998)}]{Goldie}%
  \BibitemOpen
  \bibfield  {author} {\bibinfo {author} {\bibfnamefont {C.~M.}\ \bibnamefont
  {Goldie}}\ and\ \bibinfo {author} {\bibfnamefont {C.}~\bibnamefont
  {Kl{\"u}ppelberg}},\ }in\ \href@noop {} {\emph {\bibinfo {booktitle} {A
  practical guide to heavy tails: statistical techniques and applications}}}\
  (\bibinfo {year} {1998})\ pp.\ \bibinfo {pages} {435--459}\BibitemShut
  {NoStop}%
\bibitem [{\citenamefont {Dorogovtsev}\ and\ \citenamefont
  {Mendes}(2002)}]{Dorogovtsev}%
  \BibitemOpen
  \bibfield  {author} {\bibinfo {author} {\bibfnamefont {S.}~\bibnamefont
  {Dorogovtsev}}\ and\ \bibinfo {author} {\bibfnamefont {J.}~\bibnamefont
  {Mendes}},\ }\href@noop {} {\bibfield  {journal} {\bibinfo  {journal} {Adv.
  Phys.}\ }\textbf {\bibinfo {volume} {51}},\ \bibinfo {pages} {1079} (\bibinfo
  {year} {2002})}\BibitemShut {NoStop}%
\bibitem [{\citenamefont {Bollob{\'a}s}\ \emph {et~al.}(2001)\citenamefont
  {Bollob{\'a}s}, \citenamefont {Riordan}, \citenamefont {Spencer},\ and\
  \citenamefont {Tusn{\'a}dy}}]{Bollobas}%
  \BibitemOpen
  \bibfield  {author} {\bibinfo {author} {\bibfnamefont {B.}~\bibnamefont
  {Bollob{\'a}s}}, \bibinfo {author} {\bibfnamefont {O.}~\bibnamefont
  {Riordan}}, \bibinfo {author} {\bibfnamefont {J.}~\bibnamefont {Spencer}}, \
  and\ \bibinfo {author} {\bibfnamefont {G.}~\bibnamefont {Tusn{\'a}dy}},\
  }\href@noop {} {\bibfield  {journal} {\bibinfo  {journal} {Random Structures
  and Algorithms}\ }\textbf {\bibinfo {volume} {18}},\ \bibinfo {pages} {279}
  (\bibinfo {year} {2001})}\BibitemShut {NoStop}%
\bibitem [{\citenamefont {Waclaw}\ and\ \citenamefont
  {Sokolov}(2007)}]{Waclaw}%
  \BibitemOpen
  \bibfield  {author} {\bibinfo {author} {\bibfnamefont {B.}~\bibnamefont
  {Waclaw}}\ and\ \bibinfo {author} {\bibfnamefont {I.~M.}\ \bibnamefont
  {Sokolov}},\ }\href@noop {} {\bibfield  {journal} {\bibinfo  {journal}
  {Physical Review E}\ }\textbf {\bibinfo {volume} {75}},\ \bibinfo {pages}
  {056114} (\bibinfo {year} {2007})}\BibitemShut {NoStop}%
\bibitem [{\citenamefont {Kesten}(1980)}]{squarepercolation}%
  \BibitemOpen
  \bibfield  {author} {\bibinfo {author} {\bibfnamefont {H.}~\bibnamefont
  {Kesten}},\ }\href@noop {} {\bibfield  {journal} {\bibinfo  {journal} {Comm.
  Math. Phys.}\ }\textbf {\bibinfo {volume} {74}},\ \bibinfo {pages} {41}
  (\bibinfo {year} {1980})}\BibitemShut {NoStop}%
\bibitem [{\citenamefont {Clauset}\ \emph {et~al.}()\citenamefont {Clauset},
  \citenamefont {Tucker},\ and\ \citenamefont {Sainz}}]{ICON}%
  \BibitemOpen
  \bibfield  {author} {\bibinfo {author} {\bibfnamefont {A.}~\bibnamefont
  {Clauset}}, \bibinfo {author} {\bibfnamefont {E.}~\bibnamefont {Tucker}}, \
  and\ \bibinfo {author} {\bibfnamefont {M.}~\bibnamefont {Sainz}},\
  }\href@noop {} {\enquote {\bibinfo {title} {{The Colorado Index of Complex
  Networks}},}\ }\bibinfo {howpublished}
  {\url{https://icon.colorado.edu}}\BibitemShut {NoStop}%
\bibitem [{\citenamefont {Schultes}()}]{USRoads}%
  \BibitemOpen
  \bibfield  {author} {\bibinfo {author} {\bibfnamefont {D.}~\bibnamefont
  {Schultes}},\ }\href@noop {} {\enquote {\bibinfo {title} {{United States Road
  Networks} {(TIGER/Line)}},}\ }\bibinfo {howpublished}
  {\url{http://www.dis.uniroma1.it/challenge9/data/tiger/}},\ \bibinfo {note}
  {{October 2005}}\BibitemShut {NoStop}%
\bibitem [{\citenamefont {Lee}\ \emph {et~al.}(2017)\citenamefont {Lee},
  \citenamefont {Fricker},\ and\ \citenamefont {Porter}}]{fungal}%
  \BibitemOpen
  \bibfield  {author} {\bibinfo {author} {\bibfnamefont {S.}~\bibnamefont
  {Lee}}, \bibinfo {author} {\bibfnamefont {M.}~\bibnamefont {Fricker}}, \ and\
  \bibinfo {author} {\bibfnamefont {M.}~\bibnamefont {Porter}},\ }\href@noop {}
  {\bibfield  {journal} {\bibinfo  {journal} {Journal of Complex Networks}\
  }\textbf {\bibinfo {volume} {5}},\ \bibinfo {pages} {145} (\bibinfo {year}
  {2017})}\BibitemShut {NoStop}%
\bibitem [{\citenamefont {Das}\ and\ \citenamefont {Yu}(2012)}]{Musculus}%
  \BibitemOpen
  \bibfield  {author} {\bibinfo {author} {\bibfnamefont {J.}~\bibnamefont
  {Das}}\ and\ \bibinfo {author} {\bibfnamefont {H.}~\bibnamefont {Yu}},\
  }\href@noop {} {\bibfield  {journal} {\bibinfo  {journal} {BMC Systems
  Biology}\ }\textbf {\bibinfo {volume} {6}},\ \bibinfo {pages} {92} (\bibinfo
  {year} {2012})}\BibitemShut {NoStop}%
\bibitem [{\citenamefont {Seierstad}\ and\ \citenamefont
  {Opsahl}(2011)}]{NorwegianBOD}%
  \BibitemOpen
  \bibfield  {author} {\bibinfo {author} {\bibfnamefont {C.}~\bibnamefont
  {Seierstad}}\ and\ \bibinfo {author} {\bibfnamefont {T.}~\bibnamefont
  {Opsahl}},\ }\href@noop {} {\bibfield  {journal} {\bibinfo  {journal}
  {Scandanavian Journal of Management}\ }\textbf {\bibinfo {volume} {27}},\
  \bibinfo {pages} {44} (\bibinfo {year} {2011})}\BibitemShut {NoStop}%
\bibitem [{\citenamefont {Welford}(1962)}]{Welford}%
  \BibitemOpen
  \bibfield  {author} {\bibinfo {author} {\bibfnamefont {B.}~\bibnamefont
  {Welford}},\ }\href@noop {} {\bibfield  {journal} {\bibinfo  {journal}
  {Technometrics}\ }\textbf {\bibinfo {volume} {4}},\ \bibinfo {pages} {419}
  (\bibinfo {year} {1962})}\BibitemShut {NoStop}%
\end{thebibliography}%

\end{document}